# Where the Rubber Meets the Sky:
# Bridging the Gap between Databases and Science

Jim Gray

Alex Szalay

# Where the Rubber Meets the Sky: Bridging the Gap between Databases and Science

Jim Gray
Microsoft Research
Gray@Microsoft.com

Alex Szalay
Johns Hopkins University
Szalay@pha.jhu.edu

**Abstract:** *Scientists in all domains face a data avalanche – both from better instruments and from improved simulations. We believe that computer science tools and computer scientists are in a position to help all the sciences by building tools and developing techniques to manage, analyze, and visualize peta-scale scientific information. This article is summarizes our experiences over the last seven years trying to bridge the gap between database technology and the needs of the astronomy community in building the World-Wide Telescope.*

## 1. Introduction

If you are reading this you are probably a "database person", and have wondered why our "stuff" is widely used to manage information in commerce and government but seems to not be used by our colleagues in the sciences. In particular our physics, chemistry, biology, geology, and oceanography colleagues often tell us: "I tried to use databases in my project, but they were just too [slow | hard-to-use | expensive | complex ]. So, I use files." Indeed, even our computer science colleagues typically manage their experimental data without using database tools.

What's wrong with our database tools? What are we doing wrong? We have been trying to answer those two questions in the context of astronomy data – that is, we have been working for the last seven years to try to bridge the gap between database products and the community of professional astronomers.

In the process we have come to realize that science is being transformed – most scientific branches are becoming data-rich. To deal with the data avalanche, each discipline is growing a data analysis arm that collects large datasets representing mankind's knowledge of the field. These datasets are augmented with applications that encapsulated mankind's understanding of the data and our understanding of that branch of science. Each field is being objectified – it is building an object model that represents the facts and building methods that allow easy selection and recombination of the facts. This represents one of the most exciting branches of each of these sciences – and database and data mining technologies are the heart of each of these developments.

Put more succinctly: each science was empirical at first and then grew an analytic branch. Recently the analytic branches have spawned a computational branch that simulates the analytic models. So the X department of almost all universities has a computational-X faculty (e.g.: computational-linguistics, computational-genomics, computational-chemistry, etc.). These computational scientists are adding to the data avalanche. Now we are seeing the emergence of X-info departments to deal with the avalanche (e.g., bio-informatics). The X-info community is trying to put all the data and literature on the public internet, federate it all, and provide tools that make it easy for domain scientists to find information, and make it easy for scientists to add their data and findings to the corpus. This will redefine the notion of "publishing" and has deep implications for how the world's scientific enterprise is organized.

The vision is *eScience* in which the unified experimental, theoretical, and simulation data and literature are at your fingertips. You can explore all the world's scientific data and literature looking for patterns and anomalies. Tools that encapsulate the best statistical algorithms, the best machine learning algorithms and the best data visualization are at your disposal to make you more productive and a better scientist. When you want to publish your findings, all your work has been recorded, so that others can follow your arguments and extend your work. That's the vision; but, the reality is far from that vision.

Historically, scientists gathered and analyzed their own data. But technology has created functional specialization where some scientists gather or generate data, and others analyze it. Technology allows us to easily capture vast amounts of empirical data and to generate vast amounts of simulated data. Technology also allows us to store these bytes almost indefinitely. But there are few tools to organize scientific data for easy access and query, few tools to curate the data, and few tools to federate science archives.

Domain scientists, notably NCBI (ncbi.nih.gov) and the World Wide Telescope (WWT, ivoa.net), are making heroic efforts to address these information management problems. But it is a generic problem that cuts across all scientific disciplines. A coordinated effort by the computer science community to build generic tools could help all the sciences. Our current database products are a start, but much more is needed.

We have been participating directly in the WWT effort. This is a personal and anecdotal account of what we learned working with the astronomers and advice to others on how to do similar things for astronomy or other disciplines.

## 2. Science is Changing

For each science discipline X, X-info and comp-X are controversial. There is a long-tradition of using primary sources -- gathering and analyzing your own data rather than using second-hand data or synthetic data. The traditional primary-data approach is still valid, but scientists increasingly need to compare their data to large shared databases (e.g. is this DNA sequence already in Genbank? If so, what is known about it?) In some cases advances come from analyzing existing data sources in new ways – the pentaquark was found in the archives once theoreticians told us what to look for.

Traditionally, data analysis has followed an *ftp-grep model*. In the ftp-grep model, the scientist first gets a copy of the relevant data from an archive: that is she uses ftp (file transfer protocol) to copy the relevant data sets from a science archive on the internet to a local computational server. The next step is that the scientist uses an analysis package to scan through the data looking for patterns. We use the name grep (generalized regular expressions) to describe this process. In fact the analysis program is often a "dusty deck" written long ago in Fortran invoked with a scripting language like Python.

The ftp-grep data analysis approach is breaking down as datasets grow to terabyte and petabyte scale. First, although you can ftp a megabyte in a second, it takes days or years to ftp a terabyte to your local server – and most local servers do not have spare terabytes of space. But datasets are growing to the petabyte range. So, scientists must to learn to move the questions-and-answers across the internet, and do the data analysis near the data whenever possible.

Section 4 explores this data-archive and portal architecture in more depth. But here, we want to make the point that the skills and the workflow of next generation science will be different. The new breed will need new tools and need the skills to use them.

Traditionally science data was included in tables and graphs that were part of the science article. This metaphor is breaking down or has already broken down – the data is just too large to be printed. Rather the data is placed in a project website (like [BaBar](#)) or is deposited with a science archive like [Genbank](#) or [Astrophysics Data System](#). So, the concept of publishing has morphed so that the "paper" is about three things: (1) "how to read the data", (2) "here is what we found", and (3) "here are some things we are going to look for next in the data and things you might want to look at."

In this new world, the data is published on the Internet and the literature mostly summarizes the data, is a guide to it, and presents some interesting conclusions. Not all science is like this, but the trend is definitely increasing.

Even more interesting to us is the fact that new results come from new combinations of existing data. It appears that there is a "Metcalfe's law" for datasets: the utility of N independent data sets seems to increase super-linearly. One can make $N(N-1) \sim N^2$ comparisons among the datasets and so the utility is approximately $N^2$.

The increased utility of being able to combine multiple independent datasets is one of the drivers for the digital libraries being built by the various science disciplines. The biology community has impressive portals in [Entrez](#), [DDJB](#), and [Esembl](#) – they integrate all the data and the literature. They are adding new datasets rapidly. They represent an architecture where one group manages both the data and literature all in one location. An alterative approach, typified by SkyQuery.Net lets the data reside at the data sources and a web portal ([SkyQuery.Net](#) in this case) federates them via web services.

We discuss these architectural alternatives in Section 4. The point here is that the World Wide Digital Library for each discipline X is quickly evolving – mostly driven by the domain scientists. With time we expect these many libraries to grow together to form one giant library. The architecture for these federated libraries and for the federation of libraries stands as a challenge for the computer science field. If we fail in our responsibility to address that challenge, the domain scientists will certainly come up with solutions and we computer scientists will have missed an opportunity to do something really valuable.

One problem the large science experiments face is that software is an out-of-control expense. They budget 25% or so for software and end up paying a lot more. The extra software costs are often hidden in other parts of the project – the instrument control system software may be hidden in the instrument budget. The data pipeline software budget may be hidden in some of the science budgets. But there is no easy way to hide the cost of the archive in some other budget. So the archive software often does not get written at all. In this scenario, the data is deposited in a poorly documented ftp-server and is never connected to an easy public access.

The funding agencies want scientists to publish their data, but the agencies cannot afford to pay for the data archives. One thing we computer scientists could do is to make it much easier and less expensive to publish scientific data.

## 3. Rules of Engagement – Working With X

How can you, a computer scientist, engage with a domain scientist (or group)? You might try going native – switch from computer science to bio or eco or geo or X science – stop reading all those old CS journals and give up going to all those computer science conferences.

Certainly, we have seen going native "work" but it has two shortcomings (1) you are giving up a lot, (2) you are unlikely to be able to help domains Y and Z who need you just as much as X. So, our advice is to stick to your expertise. Keep your office, your CS appointment and your magazine subscriptions; but, start a collaboration with some scientists and over time broaden that collaboration to work cross multiple disciplines.

So, how do you engage? Well, first you have to find someone in domain X who is desperate for help. Frankly, most scientists are quite happy with the status quo. They are collecting their data; they are publishing their papers; and they are generally working the way they and their colleagues have worked for decades if not centuries. All this stuff about a data avalanche is not really their problem. There is however a small group of pioneers in most fields who are absolutely desperate for help. They are pioneering a new instrument, or have a new simulator that is indeed producing the data avalanche. The answer is somewhere inside the terabytes if they could only get the data processed, organized, and then be able to query it in finite time.

You will recognize these people when you meet them – they are the ones with the jobs that take weeks or months to run their Python scripts. Their delay from question to answer is days or weeks. They are the ones who are doing batch processing on their data. They envy people who have interactive access to data and envy people who can explore their data. They are the ones who are desperate for help. Desperate enough to even spend time on a long-shot like working with you.

OK, so suppose you locate a desperate person. How can you help? How can you communicate? Well, first you have to learn a bit of their language. This generally involves mastering the introductory text for that domain – it is painful and you can skip this step if you are in a hurry, but you will end up doing this work in any case. Doing it early is the most efficient way. In parallel you have to form a working relationship with the domain experts (scientists.) You need to put in enough face time so that they are not surprised to see you. This goes hand-in hand with developing a common language. The converse of this, the domain scientists you are working with need to explore some of the things done in computer science and in other disciplines so that they have a sense of what is possible and what is almost possible.

Once you have established communication, you need to find problems that leverage your skills. This requires you do some ethnography of Domain X. Do a time-and-motion study of how the scientists are currently working, and look for some low-hanging-fruit; simple tools you could build or deploy that would be a major productivity boost. You are looking for a project that will take a few weeks or months of your time, and that would save them years of their time.

To give a specific example, in the context of the Sloan Digital Sky Survey data system, we developed a set of 20 questions each of which would have taken a week or more of programming and a week of execution with the then-existing tools. We also developed 12 visualization tasks (scenarios) that typify the kind of data visualization that scientists want to do with the Sloan Digital Sky Survey data [Szalay00].

These 20 questions then became the requirements document for the design of the SDSS Archive server. It had to be easy (less than an hour) to ask each of these questions and the answers had to come back within minutes. This was a contract that both sides (the Astronomers and the Computer Scientists) could understand.

Since that time, we have seen this 20-questions approach applied to many other branches in astronomy. Typically the new group adopts about half of the 20 questions from the original set but discards about half and adds ten more that are unique to that branch or instrument.

Our rough observation is that scientists spend a lot of time loading, scrubbing, and conditioning data. Once they have the data in acceptable form, they spend their time looking for haystacks and for needles in haystacks – that is they spend their time looking for global properties and global trends and patterns (the haystacks), or they spend their time looking for anomalies (needles in the haystacks). Computer scientists can help automate the data handling processes, they can help with the data analysis, and they can help with the scientists ask questions and understand answers.

It seems Computer Scientists can be roughly divided into three groups: (1) Data Visualization experts who are good at capturing user's questions and good at visualizing answers in novel ways. (2) Data Analysis experts who are skilled at statistical analysis, data mining algorithms, and machine learning algorithms, or just algorithms in general. (3) Software Systems people who are good at data management or systems management or workflow management. The domain scientists will likely not understand these broad distinctions, but computer scientists know that these three communities are largely segregated – they each have their own conference and journals.

Given these observations, it seems obvious that the Visualization and Analysis computer scientists can help improve existing systems, but that the logical progression is for the systems people to build a "rationalized base" and then for the Algorithms and Visualization people to build new tools atop that base. So, we have been working away to build both individual astronomy archives and to federate that as described in the next sections. We are certainly not done building the base, but we think we are ready to accommodate any Analysis or Visualization experts who want to apply their techniques to this corpus.

A rough chronology of our collaboration over the last seven years is that the twenty queries suggested a logical and physical database design – and also showed the need for a spatial data access method.  About 25% of the queries involved spatial clustering of objects.   Addressing these issues took a year or two and naturally led to an "outreach website ([SkyServer.SDSS.org](SkyServer.SDSS.org)) that provides both public access to the data and is also the basis for online astronomy education for pre-college students. There are now 150 hours of online astronomy instruction in several languages that teach both astronomy and teach tools and skills to analyze a large online scientific database.  This education component comprises more than 10% of the traffic on the SkyServer website.

At this point, data loading had become an onerous task – the data kept coming and being reprocessed.  So, we automated the loading process with a workflow system built atop SQLserver and its Data Transformation Service (DTS.) With the SkyServer in place we began exploring architectures for the World Wide Telescope (the Virtual Observatory) federating all the world's astronomy archives.

A pixel server (image cutout) web-service was the first experiment with web services.  It led to the SkyQuery (SkyQuery.Net) architecture.

In parallel, several other groups did their own 20-questions exercise and cloned the basic SkyServer database design and spatial access methods for their archives.  Recently we have been refining and extending our spatial access techniques , redoing our design in light of  the revolution in database architectures (the unification of data and methods), and also preparing for the next-generation surveys which will be capturing a thousand times more data.

In summary, it has been a very productive collaboration.  It has in part answered the question: "Why don't they use our stuff?"  The answer is indeed that databases are hard to use. Paradoxically, it is difficult to get data into them – data scrubbing and loading is labor intensive.  Organizing data is challenging.  Once the data is loaded, cognoscenti can do wonders, but "normals" do not have the data analysis skills to pose queries and to visualize answers.  As the community evolves, and as some scientists master these skills, they communicate successful patterns to their colleagues – that is how scientific techniques evolve.  It is an organic process.

In closing this section on "how to collaborate" we confess that we are very data centric. Computer scientists might contribute to domain X by bringing new ways to think about dynamic systems or new notations for dynamic systems. Differential equations are the standard way to describe dynamics 00 but they are not particularly good at describing discrete event systems.  Computer scientists have struggled with and modeled discrete event system with tools like simulation languages, process algebras, and temporal logics. It is quite possible that these new ways of describing discrete events could provide breakthroughs in process-oriented sciences as diverse as ecology, metabolic-pathways, or social interactions.

## 4.   Architecture: Services and Portals.

The broad architecture of the World Wide Telescope and of other digital libraries has emerged over the last decade.  They will be federations of data servers, each providing access to its data resources.   The federation will be unified by one or more portals, typically collocated with one of the mega-servers.  Each portal provides high-level tools to access the "local" archive and also to correlate that data with the other data sources in the internet and with the scientific literature.

This pattern is emerging in other disciplines – for example [Entrez](Entrez), [DDJB](DDJB), and [Esembl](Esembl)  show the same pattern in the bio-informatics domain. A mega-archive co-located with a portal providing tools that integrate and federate that archive with the other important archives.

The Service-Portal architecture requires that data from different data services be interchangeable – that there be standard terms and concepts that apply across all the archives.   This requires the *objectification* of the discipline – defining an object model that encapsulates facts and represents them in standard ways.   The astronomy community has embraced the UCD methodology as a core vocabulary for units and facts [UCD].  From a database perspective – this is the schema.

Built atop this, the astronomers have defined a few web services that, given a query, return the answer.  These services are defined in terms of XML Schema and SOAP messages.   The first service returned all objects within a circle (cone) [Cone]; the second returned an image of an area of the sky [SIAP]. There is also a spectrogram service.  A dozen generic services have been defined so far [VO-services].

The portals build on top of these simple services providing ways to combine data from multiple archives.   This is nascent now – there are only a few portals, but each of them is evolving rapidly.

One unanticipated benefit of developing the SkyServer architecture and placing it in the public domain is that it has been copied about ten times by other archives.   The server has a "spine schema consisting of the spatial data organization and the general organization of optical and spectroscopic data.  Other surveys have discarded the SDSS-specific attributes and replaced them with their own.  The result is that the Royal Observatory of Edinburgh, Space Telescope Institute, Cornell, Caltech, and NOAO have each been able to leverage this software base.  In general, a survey can be converted to the spine schema in a matter of hours if it is small (gigabytes) or a matter of days if it is large (terabytes).

The common spine schema and web services interface has also enabled us and others to add many datasets to the SkyQuery interface. Once the data is online, the web services can be defined and it can be registered as a member of the SkyQuery federation in a few hours [VO Registry].

The portal itself has evolved in two un-anticipated ways: (1) users can define personal databases that represent their "workspace" on the server, and (2) users can submit long-running queries to a batch system that schedules these jobs and deposits the answers into the user's personal database. Both these features are described in [Nieto-Santisteban 04] and are briefly sketched here.

Users routinely perform data analysis in steps – using the results of previous steps in subsequent analysis. The often start the analysis with datasets that they have developed over the years. To facilitate this, the portal lets a user define a database that is private to that user. The user can create tables in that database and upload private data to it. The user can deposit query answers to existing or new tables in this database, and can use tables from this database in subsequent queries. Access control is based on username and password (better security mechanisms are being discussed). The database owner can grant read or write authority to the personal database. This is convenient for collaborators who share a workspace and is a convenient way to publish data to your colleagues. The databases are currently limited to 1GB, and so far no one has asked for more – but we see no reason to limit the databases to such a small size in the future.

The portal has limits on how much processing a query can consume (elapsed time) and how large the outputs can be. For the public site the limits are 90 seconds elapsed and 1,000 rows of output. For the collaboration site the limits are 90 minutes elapsed and 500,000 rows of output. Both categories occasionally need to exceed these limits, and it is often the case that the user does not want to wait for the job to finish. This obviously suggests a batch job scheduler. Users can submit jobs to this scheduler which then runs them in the background. If a job exceeds its quota, there is an option to double the quota and re-run the job (up to some limit). There is also a simple web interface to check on the status of jobs. The easiest way to understand this system is to just try it: create a personal database and run some sample queries [CAS].

## 5. Was It Worth the Effort?

After all this work, what are the benefits? Are the scientists better able to access their data? What breakthroughs has this enabled? At best, we can answer these questions with anecdotal evidence.

The SkyServer website has served about 60M web hits. More than 10% of that traffic is to the educational part of the site – so the "outreach" has been very successful. The collaboration SkyServer website is currently processing more than 100,000 SQL queries per month, the public site processes many more and the CAS Job system is processing nearly 10,000 jobs per month. So, the system *is* being used. The astronomers have wrapped the web services in Emacs and Python scripts. Many use them as a matter of course in their work.

There are two side-by-side comparisons of the files-vs-database approach. One compared a "Grid-Condor" design for finding galaxy clusters to a database-index-parallelism approach. The DB did the job better, simpler, and 50x faster [Nieto-Santisteban 04a]. In a second experiment, a colleague had written a program to find near-earth objects (asteroids). The program was about 20 pages of code and ran for three days. We wrote a 50-line SQL program in a few hours that got the same answer in about 30 minutes. When we added an index to support the query, it ran in 3 minutes. When questions and answers can be done in minutes it changes the kinds of questions you ask – this seems to us to be an example of the breakthrough we are talking about.

The SDSS is a side-by-side optical and spectroscopic survey. Spectrograms give precise redshifts and hence approximate distances based on Hubble's constant. We use object spectra as a "training set" to estimate even more galaxy distances based on photometry alone – the so called *photometric redshifts*. We found that one category of objects was significantly under-sampled by the spectroscopic survey and so our photometric redshift estimator was less sensitive than it might be. The code to schedule and "drill" the fiber plate to take spectra was huge and poorly understood – so the astronomers could not change it easily to address this custom problem. In an evening we wrote the queries to select a set of observations and generated a drilling list (holes in an observing plate). The observation was scheduled and a few months later we got the data from the observation. Within hours the data was loaded into the database and the redshift estimator was recalibrated with the larger dynamic range and substantially improved accuracy.

Our colleagues have much better access to the SDSS data than they had to any previous survey. Not just that, having a database has improved data quality. Foreign keys, by having integrity constraints and by doing integrity checks in the data loading process have uncovered a myriad of problems which have fed back into the pipeline.

The SkyServer embeds the data documentation in the DDL (using the UCD notation and terms [UCD].) A tool scans this documentation and automatically generates online documentation for the website. So, once the astronomer has defined the DDL, he can use the automated loading tools (DTS) to load the database, and the website comes up with access to the data and with online documentation – immediately. This has been a boon to several projects – saving months if not years of development.

In summary, we think that astronomers are finding that this is the easiest way to publish their observational data. It could be better, but it is a substantial advance over what existed. Now the challenge is to make the process even more automatic, deal 1,000x more data volumes, and provide the missing visualization and analysis tools.

## 6. Some Lessons Learned From the WWT

Astronomy is a good example of the data avalanche. It is becoming a data-rich science. The computational-Astronomers are riding the Moore's Law curve, producing larger and larger datasets each year. The observational astronomers are also riding Moore's law, deploying denser CCDs and more instruments each year. The total data volume seems to be doubling every year.

We have learned some interesting lessons from managing the Sloan Digital Sky Survey dataset. It is about 40 TB of raw image data which places it somewhere between Genbank (50GB) and BaBar (1PB) in the spectrum of scientific datasets.

**Data Inflation Paradox:** You might guess that the derived data products, the processed astronomy data, would be 10% of the size of the raw instrument data. After all, the night sky is mostly black. Your intuition is approximately correct. The *Data Inflation Paradox* is that after a few years the derived data products form most of the data volume.

NASA has a useful classification for data products: Level 0 is the raw instrument data, Level 1 is the calibrated data, and Level 2 are derived data products built from Level 1 datasets. As time passes, the Level 0 data arrives at an approximately constant rate. But, a growing number of Level 1 and higher data products are derived from the raw data – star catalogs, QSO catalogs, cluster catalogs, materialized views that are particularly convenient. The derived products are indeed much smaller – but they are also much more numerous.

Versions are a second data inflation driver. The derived data products are produced by software. Software comes in versions. As the software improves, the entire Level 0 archive is reprocessed to produce a new edition of the data.

Once this edition is published, it becomes part of the science record and must remain published (unchanged) forever. Suppose you have an archive growing at one terabyte per year at level 0 with a derived data product that is 10x smaller. In year 1, the data product is 10% of the storage. In year 5 you have five terabytes of level 0 data and .1+.2+.3+.4+.5 = 1.5 TB of level 1 data. Over the *N*-year life of the project, the level 1 data volume grows as $N^2$ while the level 0 data grows as *N*. Even if there is only one data product, the derived data volume will eventually overwhelm the Level 0 data volume. But, there are many Level 1 products – and their number grows with time.

**The 5-Year peak**: When the archive starts, data arrives at a certain rate, let's say 10TB per year. After 5 years, you will have 50 TB of Level 0 data and perhaps 50 TB more of Level 1 data products. In the mean time, Moore's law will have made storage and processing about ten times less expensive. But in addition, you need to replace equipment that is more than 3 years old. So, the cash flow is a fairly complex combination of these three forces: data inflation, Moore's law, and depreciation. Figure 1 plots these three forces. Ignoring data inflation, the peak is at 2.5 years, but with data inflation it is at 5 years.

**Overpower by 6x**: However much processing and storage you think you need, you need six times as much. The first factor of two is easy to explain: if you store data, you need multiple copies just to be safe. If you do some computation, you should plan to use 50% of the processing power because things inevitably break, and mistakes are inevitably made. So, it is better to build in a factor of 2.

What about that other factor of 3? Well, for storage, you need to have the "production version" and the "next-production version" As you are building the next version you often find yourself sorting the data or indexing it or doing some operation that doubles or triples the size of the data (old version, intermediate version, target version.) You can generally find a way round this, by being careful (and lucky), but it is generally a lot more economical to optimize people's time rather than computer hardware. So, that is where the factor of 3 comes from.

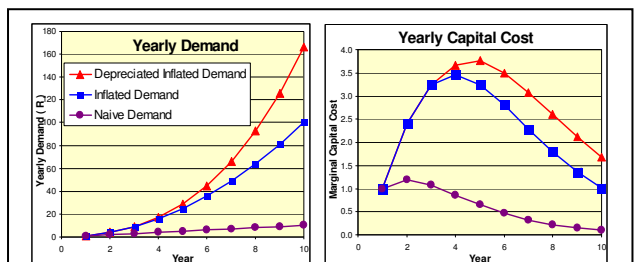

Figure 1: Archive expense (yearly outlay) for an archive where equipment depreciates over 3 years, where Moore's law improves prices by 60% per year, and where data arrives at a constant rate, where the data inflation of ten data products is quadratic.

For example, when you are reprocessing all the old stuff and you are processing the current stuff, it is important to have a substantial processing, IO, and storage over-capacity so that you can redo the work before it is time to redo the redo.

**Automate Data Loading:** A substantial part of the project's energy will go into data loading. The database will be reloaded several times – perhaps once or twice a year. So it is essential that the load process be automatic and that there be automatic ways to detect and reconcile data quality errors. People told us this before we started – but we did not believe them. They were right. We invested in this technology late and so wasted a lot of time. At this point, data loading is fairly automatic.

**Performance Regression Tests**: As the database evolves, and grows, it is essential to have a regression test. The ones we settled on were (1) we should be able to reload the entire database in a week, and (2) none of the 20 queries should take more than 10 minutes to run. Every major release causes us to fix one or another problem and to buy much better hardware in order to meet these performance goals.

**Data Pyramid**: It is very convenient to have a 1GB, 10 GB, and 100 GB subsets of the database. The smallest database fits on a laptop. It facilitates quick experiments on one or another design issue. The larger databases of course require more equipment to support them, but each has its uses.

One particularly useful aspect of the 1GB DB and the corresponding web server is that you can give it to someone as a "double-click-to-install" application. OK, now they have a fully functional personal SkyServer. Now they can throw away the SDSS data and replace it with their data. This has been a very easy way for people to get started with the code.

**Capture the Bits**: The Sloan Digital Sky Survey is a federation of more than 25 institutions and more than 200 scientists. Much of the system's design has been done with teleconferences and email and source-code-control systems and a bug-tracking database. Today, you can answer almost any question by asking the experts. But the time is coming when the experts will have forgotten or will no longer be with us. So, it is important to curate both the data and all the discussions about how it was gathered and processed. We wish we had the resources to carefully curate all the SDSS design decisions and software. We do not. Short of that, what we are doing is capturing all the bits we can – all the eMail logs, all the software versions, all the bug logs, all the meeting minutes …everything we can lay our hands on. This is all going into a big repository. Our hope is that future software will be able to organize and mine this information.

## 7. Summary

We began by commenting that science is changing and that computer-science is in a position to help its sister sciences manage the impending data avalanche. One question we sought to answer was: "Can databases help?" The simple answer is: "Yes, if…" The astronomers for example needed a few things not part of the standard packages (spherical spatial data access, statistical functions,…), and they needed database expertise to help them bridge the gap between their problem and our technology.

The sciences also face data mining and data visualization challenges that we have not addressed here (beyond pointing out how important they are.).

This is literally a golden opportunity for Computer Science to make a contribution if we will take the time to work with our colleagues in X for almost any science X.

### Acknowledgments

The SkyServer evolved from Tom Barclay's TerraServer design. The SkyServer, SkyQuery, and many other ideas described here were developed by the team consisting of Tamas Budavari, Gyorgy Fekete, Nolan Li, Maria Nieto-Santistteban, William O'Mullane, Peter Kuntz, Don Slutz, Chris Stoughton, and Ani Thakar. A.S. was supported by grants from NSF, NASA, the W.M. Keck Foundation, and the Gordon and Betty Moore Foundation.